\documentclass[twocolumn,prb]{revtex4}

\usepackage{graphicx}    
\usepackage{dcolumn}     
\usepackage{bm}          

\begin{document}
\preprint{}

\title{Magnetotunneling spectroscopy as a probe for
pairing symmetry determination
in quasi-2D anisotropic superconductors}

\author{
Y. Tanuma$^1$, Y. Tanaka$^2$,
K. Kuroki$^3$, and S. Kashiwaya$^4$
}%
%
\affiliation{
$^1$Graduate School of Natural Science and Technology,
Okayama University, Okayama 700-8530, Japan \\
$^2$Department of Applied Physics, 
Nagoya University, Nagoya, 464-8603, Japan \\
$^3$Department of Applied Physics and
Chemistry, The University of Electro-Communications,
Chofu, Tokyo 182-8585, Japan \\
$^4$National Institute of Advanced Industrial Science
and Technology, Tsukuba, 305-8568, Japan
}
%
\date{\today}
\begin{abstract}
As a probe to determine the pairing symmetry of 
quasi-two-dimensional anisotropic superconductors, 
we propose tunneling spectroscopy in the presence of  magnetic field, 
where the magnetic field is parallel to the two dimensional 
planes and rotated.
As a case study, we apply this idea 
to the models of high-$T_{C}$ cuprates and 
organic superconductors $\kappa$-(ET)$_2X$.
The surface density of states at the Fermi energy 
exhibits a characteristic oscillation 
upon rotating the direction of the magnetic field 
due to the Doppler shift of the energy of quasiparticles. 
The surface density of states has a minimum 
when the applied magnetic field is  
parallel to the  node direction of the pair potential 
independent of the detailed shape of the Fermi surface. 
The amplitude of the  oscillation
is sensitively affected by the shape of 
Fermi surface. 
\end{abstract}
\maketitle

%
Nowadays, there are various superconductors
in which the pairing mechanism seems to be unconventional.
In order to clarify the pairing mechanism of
these superconductors, it is crucial to identify the pairing symmetry,
which is characterized by the sign change and the 
presence of nodes in the pair potential.
In order to determine the pairing symmetry,
several phase-sensitive probes have been used
\cite{SR95,Harlin,Tsuei},
among which is the tunneling spectroscopy, 
which enables us to detect the sign change in the pair potential 
as well as its nodal structure \cite{KT00,Lofwander,TK95}.
Namely, a reliable evidence of the sign change 
of the pair potential
is an observation of a zero-bias conductance peak
(ZBCP) \cite{TK95}, which is originated from Andreev bound
states (ABS) formed at the surfaces or interfaces
\cite{Buch,Hu,TK95,FRS97}.
The existence of ZBCP has been actually 
observed for several unconventional superconductors 
\cite{Buch,Hu,TK95,FRS97,GXL88,Alff,Covin,Wei,Wang,Iguchi,Mao,Laube,Walti}. 
\par
In principle, it is possible to determine the 
pairing symmetry through the tunneling spectroscopy via ABS
if one can make well oriented surfaces or interfaces  
for arbitrary orientations. 
However, in quasi-two dimensional (2D) superconductors, 
it is by no means easy to make a well oriented 
surface/interface in the direction perpendicular to the 2D planes.
%
For example, for the high-$T_{C}$ cuprates, 
it is predicted theoretically that 
ZBCP is observed most prominently for (110) 
surface/interface, and is not observed for (100) 
oriented surface/interface. \cite{TK95} 
However, in the actual experiments, 
it is not easy to prepare well controlled 
surface/interface, and only few exceptional cases 
\cite{Wei,Wang,Iguchi} have succeeded  
in discriminating  the difference
between the tunneling conductance 
of (100) and (110) oriented 
junctions. 
Moreover, it has been clarified that atomic-scale 
roughness has influence on the tunneling spectroscopy 
\cite{FRS97,Tanu1,Tanu2,Samo}.
Thus, in order to clearly determine the position of the nodes 
in the pair potential, $i.e.$, the symmetry of the pair potential,  
through tunneling spectroscopy via ABS, 
a preparation of well oriented junctions is necessary.
\par
The situation can even be more severe for 
quasi-2D organic superconductors such as 
$\kappa$-(BEDT-TTF)$_2X$ salts \cite{IYS}
[abbreviated $\kappa$-(ET)$_2X$ hereafter],
($X=\mathrm{Cu(NCS)}_2,\mathrm{Cu[N(CN)}_2]\mathrm{Br}$,
or $\mathrm{I}_3$) \cite{IYS}, which is another 
candidate for anisotropic pairing superconductors.
It is even more difficult to make 
well oriented junctions compared to the 
high-$T_{C}$ cuprates.
\par
For quasi-2D superconductors, it is much more promising 
to make films well oriented in the direction {\it parallel} to the 
planes, but the problem is that in that case, ZBCP is not obtained in 
the tunneling spectroscopy due to the absence of 
the interference of the sign change of the 
pair potential felt by the quasiparticle. 
Thus, even if there are nodes in the pair potential, 
only a V-shaped tunneling conductance is expected, 
and the direction of the nodes cannot be determined.
In order to overcome this difficulty,  
we propose in the present study tunneling spectroscopy for 
surfaces parallel to the planes in the presence of a magnetic field 
applied parallel to the planes. 
\par
In the presence of a magnetic field,
it is known that 
the energy spectrum of the quasiparticle is influenced by the 
Doppler shift \cite{FRS97,Vek99}. 
Due to this effect, specific heat and 
thermal conductivity show 
characteristic  oscillation by rotating the direction of 
the magnetic field within the 2D plane reflecting the nodal 
structure of the pair potential  
\cite{Vek99,Maki,Yu,Aubin}.
Recently, based on this idea, pairing symmetry of several 
unconventional superconductors have been discussed 
using thermal conductivity measurements
\cite{Izawa1,Izawa2,Izawa3}.
Although thermal conductivity and specific heat measurements 
are powerful ways to determine the pairing symmetry, these quantities  
are strongly influenced by phonons, so that the 
actual analysis of the experimental data, namely the subtraction of the 
phonon contributions, can be subtle. 
Thus, it is desirable to propose a 
complementary method to determine the positions of nodes. 
In this sense, tunneling spectroscopy in the presence of a magnetic field, 
{\it magnetotunneling}, is a promising method \cite{comment2002}. 
In the present paper, we apply our idea to models of 
the above mentioned quasi-2D superconductors,
namely the high-$T_{C}$ cuprates and $\kappa$-(ET)$_2X$.
\par
As regards $\kappa$-(ET)$_2X$, 
the existence of nodes of the pair potentials 
has been suggested from various experiments 
\cite{Mayaffre,Soto,Kanoda96,Nakazawa,Carrington,Pinteric}.
However, the position of nodes is still a controversial issue. 
Although early theories support 
$d_{x^2-y^2}$-wave pairing \cite{Schma,KKKM,KA99},
recent experiments  \cite{Arai,Izawa3} 
support  $d_{xy}$ pairing \cite{commentBEDT}.
Stimulated by these recent experiments, 
two of the present authors revisited this issue theoretically, 
and found  that a $d_{xy}$-like pairing can slightly dominate  
over $d_{x^2-y^2}$ pairing
when the dimerization of the BEDT-TTF molecules is not so strong
\cite{Kuro02}. 
Thus, it is intriguing to clarify 
what is expected in the magnetotunneling spectroscopy for 
each of the plausible pairing symmetries. 
\par
We calculate surface density of states (SDOS)
at the Fermi energy,
which corresponds to zero-bias tunneling conductance 
in the high barrier limit \cite{KT00}.
We find that SDOS oscillates as a function of the direction 
of the magnetic field, and  
has a broad maximum for the magnetic field applied
in the antinodal direction, 
and a minimum for the magnetic field along the node.
Although this result is similar to those obtained 
in Ref.~\onlinecite{Vek99}, where the bulk density of states 
is considered for free electrons
(namely for a round Fermi surface),
it is by no means evident from the 
beginning whether similar results
can be obtained at \textit{surfaces}
and for \textit{realistic} lattice structures
and band fillings 
(namely for realistic shapes of the Fermi surface).
\par
%
\begin{figure}[htb]
\begin{center}
\scalebox{0.55}{
\includegraphics{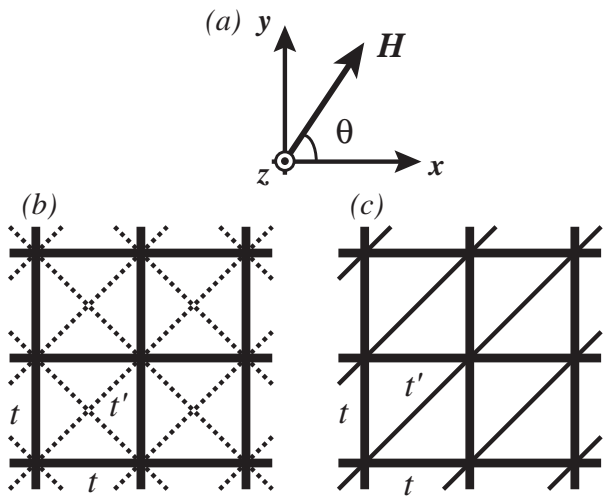}}
\caption{
(a) 
Magnetic field $H$ rotated with
the angle $\theta$ in $xy$ plane.
(b) Square lattice
and
(c) anisotropic triangular lattice
with next-nearest
neighbor hopping $t^{\prime}$.
Solid lines denote the Cooper pairs
in real space.
\label{fig:01}
}
\end{center}
\end{figure}
%
A schematic illustration of 
the models of quasi-2D superconductors 
and the direction of the magnetic field 
are shown in Fig.~\ref{fig:01},
where magnetic field is
rotated within a $xy$-plane.
Since the magnitude of the penetration depth is much 
more larger than that of coherence length in 
both superconductors, 
the vector potential can be expressed to be 
$\bm{A}(\bm{r})=
(H\lambda e^{z/\lambda}\sin \theta,-H\lambda e^{z/\lambda}\cos \theta,0)$,
where $\theta$ is the direction of
the magnetic field as defined in Fig.~\ref{fig:01}.
Thus the quasiparticle momenta $k_x$ and $k_y$
in the $x$ and $y$ directions can be chosen 
as $\tilde{k_x} = k_x +\frac{H}{\pi \xi H_{c}}\cos \theta$
and $\tilde{k_y} = k_y -\frac{H}{\pi \xi H_{c}}\sin \theta$,
where $H_{c}=\phi_{0}/(\pi^{2} \xi \lambda)$ 
with $\phi_{0}=h/(2e)$.
\par
Within the BCS mean-field theory,
in terms of the eigenenergy $E$ and the 
wave functions $u_{\bm{k}}$, $v_{\bm{k}}$,
the Bogoliubov-de Gennes equation
with the magnetic field applied  parallel to the 
$xy$-plane is given by
\begin{eqnarray}
\label{eq:BdG}
     \left (
      \begin{array}{cc}
    \xi_{\bm{k}} &
    \Delta_{\bm{k}} \\
    \Delta^{*}_{\bm{k}} &
   -\xi_{-\bm{k}} \\
      \end{array}
     \right )
     \left (
      \begin{array}{c}
            u_{\bm{k}} \\
            v_{\bm{k}} \\
      \end{array}
     \right )
=E \left (
      \begin{array}{c}
            u_{\bm{k}} \\
            v_{\bm{k}} \\
      \end{array}
     \right ).
\end{eqnarray}
\begin{eqnarray}
E_{\pm \bm{k}}=\frac{1}{2} \left [
(\xi_{\bm{k}}-\xi_{-\bm{k}})
\pm \sqrt{(\xi_{\bm{k}}+\xi_{-\bm{k}})^{2}
+4|\Delta_{\bm{k}}|^{2}}
\right ],
\end{eqnarray}
\begin{eqnarray}
|u_{\bm{k}}|^{2}=\frac{1}{2}
\left (1+\frac{\eta_{\bm{k}}}{\Gamma_{\bm{k}}} \right ),
\quad
|v_{\bm{k}}|^{2}=\frac{1}{2}
\left (1-\frac{\eta_{\bm{k}}}{\Gamma_{\bm{k}}} \right ),
\end{eqnarray}
with $\eta_{\bm{k}} = \xi_{\bm{k}}+\xi_{-\bm{k}}$
and $\Gamma_{\bm{k}} = \sqrt{\eta_{\bm{k}}^{2}
+4\Delta_{\bm{k}}^{2}}$.
\par
In the following, we discuss the $k$ dependence of 
$\xi_{\bm{k}}$ and $\Delta_{\bm{k}}$
for the two cases considered.
(i) the cuprates: We adopt the extended $t$-$J$ model
on 2D square lattice,
which is considered to describe the low-energy excitations
as a function of doping concentration $\delta$
and superexchange interaction $J$.
The reason why we use this model
is that we can systematically 
investigate the doping dependence of Fermi surface.
Since it is difficult to treat the constraint
in the $t$-$J$ model,
we apply the Gutzwiller approximation
\cite{Zhang}.
In this case, $\xi_{\bm{k}}$ and 
$\Delta_{\bm{k}}$  in Eq.(\ref{eq:BdG}) are given by
\begin{eqnarray}
\xi_{\bm{k}}&=&
-2\left ( t_{\rm eff} + J_{\rm eff}\chi \right )
\left ( \cos \tilde{k_x}a + \cos \tilde{k_y}a  \right )
\nonumber \\
&&-4t_{\rm eff}^{\prime}\cos \tilde{k_x}a \cos \tilde{k_y}a
-\mu,
\end{eqnarray}
and 
\begin{eqnarray}
\label{eq:x2-y2}
\Delta_{\bm{k}}=2\Delta_0\left (
\cos k_{x}a - \cos k_{y}a  \right ).
\end{eqnarray}
with 
$t_{\rm eff}= \frac{2\delta}{1+\delta}t$, and
$J_{\rm eff} = \frac{3}{4}\frac{4}{(1+\delta)^{2}}J$.
Here, the quantities  $\chi$,   $\mu$, and 
$\Delta_0$ are determined  self-consistently 
for each doping ratio $\delta$. 
In the following calculation, 
we fix  $J/t=0.3$ and 
$t^{\prime}/t=-0.5$ [$t^{\prime}/t=0.5$] 
as a typical values of 
hole-doped [electron-doped] cases. 
\par
%
(ii) $\kappa$-(ET)$_2X$: 
Although this material  has four molecules
per unit cell (four bands), we may reduce it to a single band model 
for the present purpose \cite{Tamura,Kino}.
The single band description is not sufficient for 
determining which pairing symmetry is more 
favorable,\cite{Kuro02} but it suffices for the argument here, where we 
discuss the tunneling spectrum for a given pairing symmetry.
In the single band description, 
the quantity $\xi_{\bm{k}}$ in Eq.(\ref{eq:BdG}) is given by
\begin{eqnarray}
\xi_{\bm{k}}&=&
-2t\left ( \cos \tilde{k_x}a + \cos \tilde{k_y}a  \right )
\nonumber \\
&&-2t^{\prime}\cos (\tilde{k_x} +\tilde{k_y})a
-\mu.
\end{eqnarray}
The values of $t$, $t^{\prime}$, and $\mu$
are chosen as to reproduce the Fermi surface
observed by Shubnikov-de Haas experiments 
\cite{Oshima}.
As for plausible pairing symmetries in $\kappa$-(ET)$_2X$,
we consider the $d_{x^2-y^2}$-wave pairing given
by Eq.(\ref{eq:x2-y2}),
and a $d_{xy}$-like pairing given by 
\begin{eqnarray}
\label{eq:xy}
\Delta_{\bm{k}}=2\Delta_0\left [
\cos k_{x}a + \cos k_{y}a - \alpha
\cos (k_{x}a +k_{y}a) \right ],
\end{eqnarray}
with $\alpha \sim 0.8$. 
These pairing symmetries have been found to 
closely compete with each other in Ref.~\onlinecite{Kuro02}.
\par
%
In order to compare our theory with 
scanning tunneling microscopy (STM) experiments,
we assume that the STM tip is metallic
with a constant density of states, 
and that the tunneling occurs only for the site nearest to the tip.
This has been shown to be valid 
through the study of tunneling conductance of unconventional 
superconductors \cite{KT00}. 
The tunneling conductance spectrum at zero-energy
is then given
at low temperatures by 
the normalized SDOS \cite{KT00},
\begin{eqnarray}
   \rho(\theta) &=&
   \frac{
   \displaystyle{
   \int ^{\infty}_{-\infty}{\rm d}\omega \rho_{\rm S}(\omega)
{\rm sech}^{2}\left ( \frac{\omega}{2k_{\rm B}T} \right )}}
   {
   \displaystyle{
   \int ^{\infty}_{-\infty}{\rm d}\omega \rho_{{\rm N}}(\omega)
{\rm sech}^{2}\left ( \frac{\omega - t}{2k_{\rm B}T} \right )}},
\\
\rho_{\rm S}(\omega) &=& \frac{1}{2}
\sum_{\bm{k}} \left \{ |u_{\bm{k}}|^{2}
\left [
 \delta(\omega - E_{\bm{k}})
+\delta(\omega - E_{-\bm{k}}) \right ]
\right .
\nonumber \\
&& \left .
+|v_{\bm{k}}|^{2} \left [
 \delta(\omega + E_{ \bm{k}})
+\delta(\omega + E_{-\bm{k}})
\right ]
\right \}.
\end{eqnarray}
Here $\rho_{\rm S}(\omega)$ denotes the 
SDOS for the superconducting state while $\rho_{\rm N}(\omega)$
the bulk density of states in the normal state.
\par
%
\begin{figure}[htb]
\begin{center}
\scalebox{0.45}{
\includegraphics{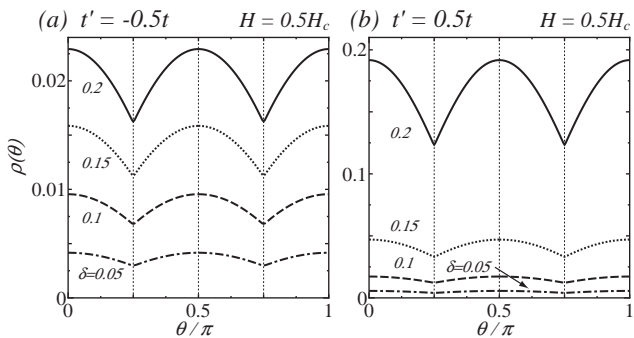}}
\caption{
The angle variation of SDOS for the model of cuprates
in $H=0.5H_c$.
(a) $t^{\prime}/t=-0.5$ and (b) $0.5$.
\label{fig:02}}
\end{center}
\end{figure}
\begin{figure}[htb]
\begin{center}
\scalebox{0.5}{
\includegraphics{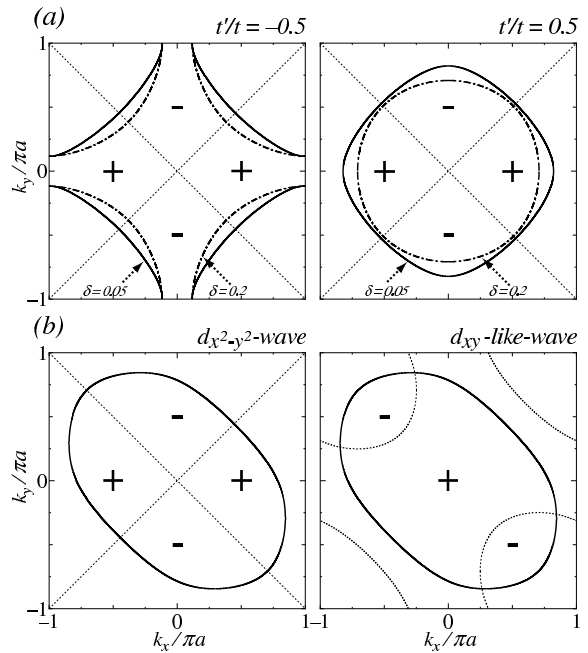}}
\caption{
The Fermi surface and $d$-wave pairings.
Dots lines represent the nodal lines.
(a) a model of cuprates
with a $d_{x^2-y^2}$-wave:
left and right panels are plotted
for $t^{\prime}/t=-0.5$ and $0.5$, respectively.
(b) a model of $\kappa$-(ET)$_2X$
for $d_{x^2-y^2}$-wave (left) and $d_{xy}$-like (right).
\label{fig:03}}
\end{center}
\end{figure}
%
First, we look into the case of the cuprates.
$\rho(\theta)$ is plotted in Fig.~\ref{fig:02}
for various $\delta$.
In the overdoped regime, 
SDOS exhibits clear fourfold oscillations, which has
a maximum for the magnetic field applied
in the antinodal direction,
and a minimum for the magnetic field along the nodes.
Thus, through the analysis of the position of 
maxima and minima of $\rho(\theta)$,
we can identify the nodal and antinodal directions.
With the increase of $\delta$, the magnitude of the 
difference between the maximum and minimum 
is enhanced. 
In the  underdoped  region, 
the difference of the magnitude 
between maximum and minimum is reduced 
reflecting on the square shape of Fermi surface
(see Fig.~\ref{fig:02}).
We now move on to the case of $\kappa$-(ET)$_2X$.
For both of $d_{x^2-y^2}$ and $d_{xy}$-like cases, 
$\rho(\theta)$ exhibits a characteristic 
oscillation reflecting the nodal 
structure and the twofold symmetry of the Fermi surface.  
For $d_{x^{2}-y^{2}}$-wave pairing, $\rho(\theta)$ has local minimum
[maximum] around  $(2n+1)\pi/4$  [$n\pi/2$] with integer $n$,   
while for $d_{xy}$-wave pairing, 
$\rho(\theta)$ has local minimum 
[maximum] around  $n\pi/2$ [$(2n+1)\pi/4$].  
As seen from this result, we can discriminate  
$d_{x^2-y^2}$-wave from $d_{xy}$. 
\par
%
\begin{figure}[htb]
\begin{center}
\scalebox{0.4}{
\includegraphics{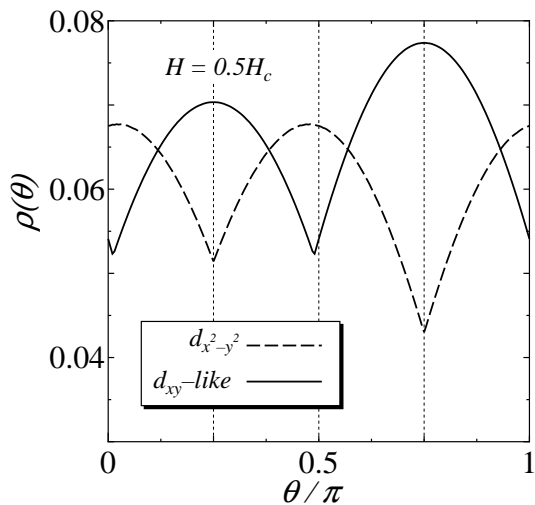}}
\caption{
The angle variation of SDOS for a model of $\kappa$-(ET)$_2X$.
\label{fig:04}}
\end{center}
\end{figure}
%
To summarize, 
we have proposed tunneling spectroscopy in the presence of a  magnetic field, 
\textit{magnetotunneling}, 
for quasi-2D anisotropic superconductors, 
where the magnetic field is 
rotated within the 2D plane. 
SDOS at the Fermi energy, 
$\rho(\theta)$,  exhibits a characteristic oscillation 
upon rotating the direction of the magnetic field. 
As case studies, we have applied this idea to the high $T_{C}$ cuprates and 
$\kappa$-(ET)$_2X$ salts.
$\rho(\theta)$ is found to take its (local) minimum value
when the applied magnetic field is  
parallel to the nodal direction of the pair potential 
{\it independent of the detailed shape of the Fermi surface}. 
The origin of the oscillation $\rho(\theta)$
is considered to be essentially the same as
that given in Ref.~\onlinecite{Vek99}, but 
we stress that it is not expected from the beginning that the 
phase of the oscillation is the same as
that in Ref.~\onlinecite{Vek99}
for arbitrary doping concentrations and/or
for various lattice structures.
In fact, we have seen that in the underdoped regime of the cuprates, 
the oscillation, although it still has the same phase as that 
in Ref.~\onlinecite{Vek99}, becomes extremely weak reflecting the 
characteristic band structure near half-filling. 
Although only two kinds of  quasi-2D superconductors 
are studied here,
the proposed \textit{magnetotunneling} spectroscopy
should serve as a strong probe to identify the pairing symmetry 
for various unconventional superconductors. 
\par
%
%
Y.T acknowledges a financial support
of Japan Society for the Promotion
of Science  for Young Scientists.
This work was in part supported by
the Core Research for
Evolutional Science and Technology (CREST)
of the Japan Science and Technology Corporation.
The computations were performed
at the Supercomputer Center of
Yukawa Institute for Theoretical Physics,
Kyoto University.
%


\begin{thebibliography}{99}
\bibitem{commentBEDT}
Throughout this paper, we refer to the pairing symmetry 
of $\kappa$-(ET)$_2X$ using the single band description
as opposed to Ref.~\protect\onlinecite{Kuro02}.
See Fig.~\ref{fig:03}(b).
\end{thebibliography}
\end{document}